\documentclass[12pt]{iopart}
\usepackage{graphicx}
\begin{document}
           \csname @twocolumnfalse\endcsname

\title{The influence of Pauli blocking effects on the properties of dense 
hydrogen}

\author{W.~Ebeling$^{1}$, D.~Blaschke$^{2,3}$, R.~Redmer$^{4}$, 
H.~Reinholz$^{4,5}$, G.~R\"opke$^{4}$}

\address{$^1$ Institute of Physics, Humboldt University, 12489 Berlin, Germany,
\\
$^2$ Institute for Theoretical Physics, University of Wroclaw, 50-204 Wroclaw, 
Poland, \\
$^3$ Bogoliubov Lab. for Theoretical Physics, JINR Dubna, 141980 Dubna,
Russia,\\
$^4$ Institute of Physics, University of Rostock, 18051 Rostock, Germany,\\
$^5$ School of Physics, University of Western Australia, Crawley, 6009 WA, Australia.}

\begin{abstract}
  We investigate the effects of Pauli blocking on the properties of hydrogen
  at high pressures, where recent experiments have shown a transition from 
  insulating behavior to metal-like conductivity. 
  Since the Pauli principle prevents multiple occupation of electron states 
  (Pauli blocking), atomic states disintegrate subsequently at high densities 
  (Mott effect).
  We calculate the energy shifts due to Pauli blocking and discuss the Mott 
  effect solving an effective Schr\"odinger equation for strongly correlated 
  systems.
  The ionization equilibrium is treated on the basis of a chemical approach.
  Results for the ionization equilibrium and the pressure in the
  region 4\,000 K $< T <$ 20\,000 K are presented.
  We show that the transition to a highly conducting state is softer than found
  in earlier work. 
  A first  order phase transition is observed at  $T<6\,450$ K, but a diffuse 
  transition appears still up to $20\,000$ K.
 \end{abstract}
\date{\today}
\section{Introduction}

The physical properties of dense hydrogen are a topic of high
interest, in particular the transition of hydrogen to a highly
conducting phase, which is considered a type of Mott transition.
Here, we will consider the effects of Pauli's exclusion principle
on high-density hydrogen plasmas. The previous studies of dense
hydrogen included several hypothetical assumptions about the
character of the high-density phase
\cite{RotesBuch,GruenesBuch,KSKBuch,Kagan,Eb84,EbRi85,Chabrier}.
While metallization of solid hydrogen near $T=0$  K has not been
clearly verified so far for pressures of up to 300 GPa
\cite{dasilva}, metal-like features have been observed in
shock-compression \cite{dasilva,weir,collins98}. 
Metal-like conductivities have been observed around 140 GPa and 3\,000 K
\cite{weir}. 
Recent experiments were able to reach that region as
well and to provide detailed information on the equation of state (EoS) 
and the conductivity in the Mbar-region \cite{fortov}. 
The transition to metal-like behavior changes drastically our present
understanding of the behavior of hydrogen at ultra-high pressures.
In the present paper, we will show that the most important effect
leading to the destruction of bound states is Pauli blocking. 
Due to the Pauli exclusion principle, the free electrons in the plasma
cannot penetrate into the interior of atoms and molecules. 
At high densities this leads to an enormous pressure acting on the
neutrals which will finally be ionized. 
The effective energy levels of hydrogen, which strongly depend on density 
and weakly on temperature, are introduced into the thermodynamic functions. 
In order to calculate the ionization/dissociation equilibrium we
minimize the free energy with respect to the composition.
Recently, we derived an expression for the free energy of dense
hydrogen $F_{\rm H}$ \cite{befjnrr,befjrr99,EbHaJRR5} in the framework of the
chemical picture and calculated the degree of ionization $\alpha$
and the degree of dissociation $\beta$ as well as the isothermal EoS, 
the hugoniots and the isentropes. 
Pauli blocking was taken into account by the concept of excluded volume,
which is based on the idea of space occupation by atoms and
molecules. 
We will show that a more fundamental approach
based on an effective Schr\"odinger equation \cite{RKKK78,RKKKZ78}, leads 
to important modifications of the earlier results.

\section{Effective Schr\"odinger equation and bound states of pairs}

\begin{figure}
 \includegraphics[width=0.5\textwidth,height=0.4\textwidth,angle=0]{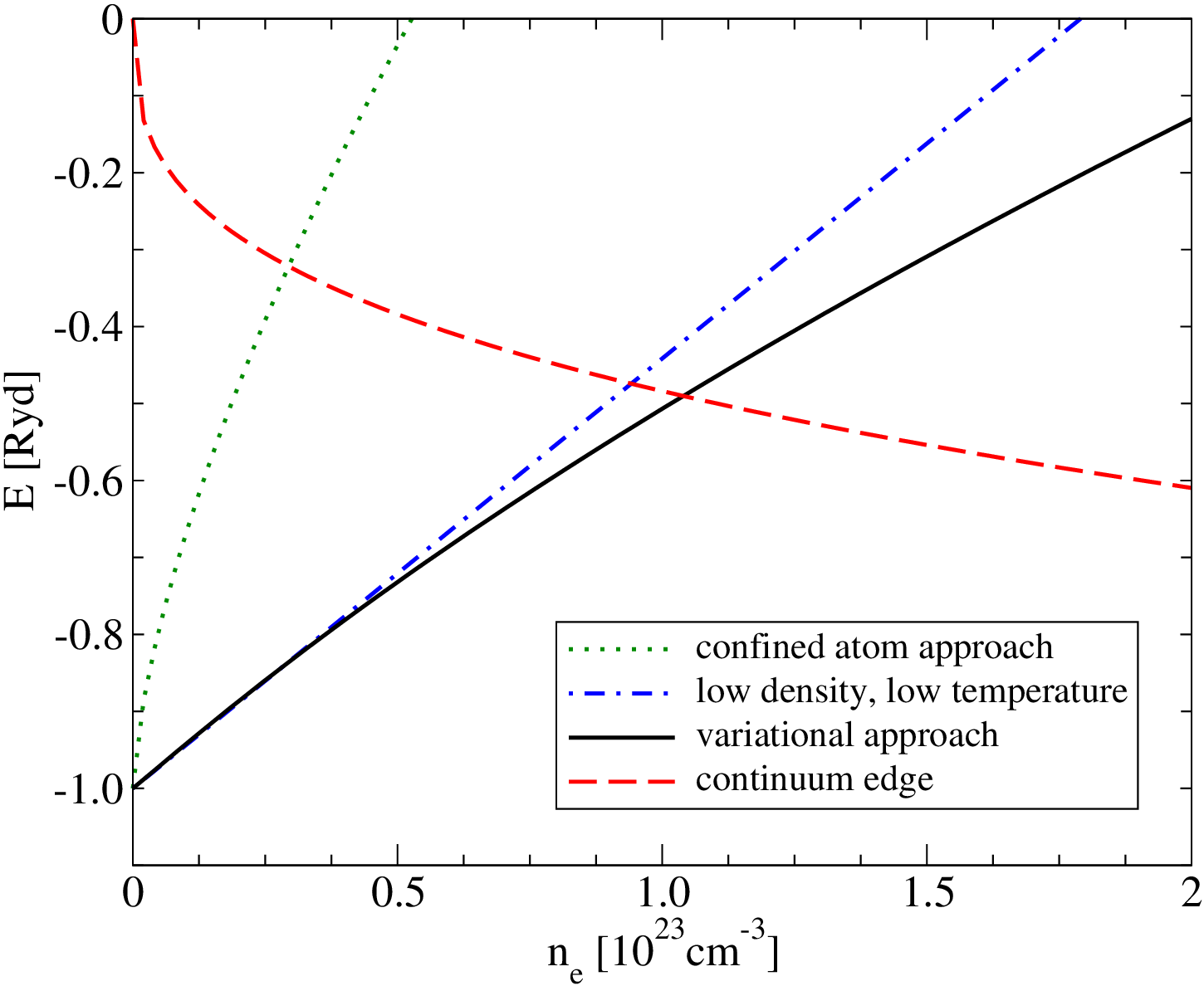} 
\caption{ \label{effener0}
 Density dependence of the effective ground state energy of hydrogen. 
 The low-temperature estimate (dash-dotted line) according to 
 Eq. (\ref{paulifock}) is compared with the confined atom approach 
 (dotted line). The solid line corresponds to the variational approach.
 We have shown also the lowering of the continuum edge (dashed line) according 
 to Eq. (\ref{contFock}).}
 \end{figure}

We focus on the interaction between atoms and free electrons and present a
microscopic treatment based on the underlying Pauli exclusion principle.
In the following we will use Rydberg units with 
$m_e = 1/2, \hbar = 1, e^2/4 \pi \epsilon_0= 2$, 
so that the binding energies of the isolated hydrogen atom are simply 
$ E^0_{n} = -1/n^2$.
Embedding the hydrogen atom in a plasma environment, the interactions with 
the medium are treated by an effective wave equation 
\cite{GruenesBuch,RKKK78,RKKKZ78}
\begin{eqnarray}
p^2 \psi_n(p)- \sum_q V(q) \psi_n(p+q)
+ \sum_q H^{\rm pl}(p,q) \psi_n(p+q)
= E_n \psi_n(p)\,,
\end{eqnarray}
where $V(q)=8\pi/q^2$ denotes the Fourier transform of the Coulomb
potential. The center of mass motion has been neglected, assuming the adiabatic limit  
$m_e/m_p \ll 1$.
In general, the plasma Hamiltonian $H^{\rm pl}(q)$ will depend also on 
$\vec P$ and on the energy, if dynamical and retardation effects are taken 
into account. 
The  plasma Hamiltonian will shift the energy eigenvalues 
$E_n = E^0_n + \Delta E_n $ and will modify the wave functions $\psi_n(p)$.
In particular, due to the plasma interaction the binding energies may merge 
with the continuum so that the bound states disappear, if the influence of 
the plasma increases with increasing density.
This dissolution of bound states is called Mott effect and has important 
consequences for the macroscopic properties of the plasma.
Let us evaluate now the mean-field energy shift of bound states writing the
effective Hamiltonian of pairs in a plasma as
\begin{eqnarray}
\label{mf1}
H^{\rm pl}(p,q) = \sum_{q'} f_e (p+q') [V(q) \delta(q') - V(q') \delta(q)] ,
\end{eqnarray}
 $f_e(p)= 1/(\exp[\beta (p^2/2m - \mu )]+1)$ is the Fermi 
distribution.
Within first order, the shift of the energy eigenvalues is obtained with the 
unperturbed wave functions $\phi_n(p)$ as
\begin{equation}
\label{pauli}
E_n-E^{0}_n = \sum_{p,p'} \phi^*_n(p) V(p' - p)  f_e(p)
\phi_n(p')- \sum_{p,p'} \phi^*_n(p)V(p' - p)  f_e(p') \phi_n(p).
\end{equation}
The first term in Eq. (\ref{pauli}) is the Pauli shift which
is due to  Pauli blocking, and can be rewritten by 
inserting the Schr\"odinger equation as
\begin{equation}
 \Delta E_n^{\rm Pauli} = \sum_{p} \phi^*_n(p) (p^2-E_n^0)  f_e(p) \phi_n(p).
\end{equation}
A simple expression is found in the low-temperature and low-density
limit, where the Fermi distribution with the normalization $\sum_p
f_e(p) = n_e/2$ is concentrated near $p=0$. 
In the zero temperature limit, we have a Fermi sphere with Fermi momentum
$p_F= (3 \pi^2 n_e)^{1/3}$. 
The energy shift  of the ground state is then
\begin{equation}
\Delta E_1^{\rm Pauli}=\frac{1}{2} n_e\,(- E_1^0)\,|\phi_1(0)|^2=32\pi n_e.
\end{equation}
At intermediate temperatures we may approximate the Fermi distribution by a 
Boltzmann distribution and find
\begin{equation} \label{paulit}
\Delta E_1^{\rm Pauli}(T) 
=  32 \pi n_e G(T/T_0) \simeq {32 \pi n_e}/[{1 + (77/ 16)\ T/ T_0}]~,
\end{equation}
where
$G(x)=\{\sqrt{x}(1+1/x)
-\sqrt{\pi}(1-x-x^2/4)[1-{\rm erf}(1/\sqrt{x})]\exp(1/x)\}/x^{7/2}$, see also \cite{EbeRed08}, 
and $T_0 = 1~{\rm Ryd}/k_{\rm B}= 157\,886$ K is the ionization temperature.
Similarly, the Fock term as the second term in Eq. (\ref{pauli}) can be 
evaluated as
\begin{equation}
\Delta E_1^{\rm Fock}=
-128 n_e~\int_0^{\infty} d p \frac{1}{(1+p^2)^4} = - 20 \pi n_e.
\label{FockB}
\end{equation}
It compensates partially the Pauli shift. Since the temperature dependence of 
the Fock shift is similar to
that of the Pauli shift we may use in first approximation the same
temperature function as in (\ref{paulit}) and  the total shift is approximated by
\begin{equation}
\label{paulifock}
\Delta E_1^{\rm Fock}+\Delta E_1^{\rm Pauli} =  12 \pi n_e G(T/T_0).
\end{equation}
The shift is shown in Fig. \ref{effener0}, dashed line, indicating
a rather steep increase of the bound state energy with density.
Due to phase space occupation, the ground state energy is shifted
and may merge with the continuum of scattering states, indicating
the dissolution of bound states. 
Considering in Eq. (\ref{mf1}) the continuum part of the spectrum describing 
scattering states, only the Fock shift contributes to the energy shift. 
The lowest energy in the continuum occurs at $p=0$ and is shifted by 
$\Delta E^{\rm Fock}(p=0) = -\sum_q V(q) f_e(q)=-4 p_F/ \pi
=- 4 (3n_e/\pi)^{1/3}$. 
However, the two-particle continuum state can only be created at the 
Fermi momentum since all states below that are occupied. 
Thus the continuum of scattering states begins at $p_F$ where we have 
in the zero temperature limit the Fock shift
\begin{equation}
\label{contFock}
\Delta E^{\rm Fock}(p_F) = - \sum_p V(\vec p - \vec p_F) f_e(p)
=- 2  p_F / \pi =- 2 (3 /\pi)^{1/3} n_e^{1/3},
\end{equation}
shown also by the dashed line in  Fig. \ref{effener0}.
Extrapolating the low-density results to higher densities, the
ground state disappears in first approximation at a density  
$n_e \simeq 0.015$. 
This corresponds to an average distance of $r_0 \simeq 2 a_B$ and is below 
the Mott criterion.
The Mott condition $r_0 \simeq a_B$ with $(4 \pi/3) n_e r_0^3 = 1$
expresses the idea that atoms are destroyed if the mean distance
of the electrons is equal or smaller than the Bohr radius. There
exist many alternative estimates of the binding energy shift.
For example, the confined atom model \cite{GruenesBuch,graboske69}
assumes that the atom is embedded into a sphere with radius $r_0$. 
In first approximation this theory gives the shift (in
Rydberg units) $\Delta E_1^{\rm ca}= \pi^2 r_0^{-2}$.
Correspondingly, the binding energy would disappear at $r_0 \simeq 3 a_B$,
i.e., already at a much smaller density (see also Fig. \ref{effener0}). 
Better estimates based on numerical solutions of the Schr\"odinger equation 
give a value of about $r_0 \simeq 2 a_B$ \cite{graboske69}. 
Our estimate is in the same region.

\section{Evaluation of the mean-field energy shift by variational approach}

\begin{figure}
 \includegraphics[width=0.5\textwidth,height=0.4\textwidth,angle=0]{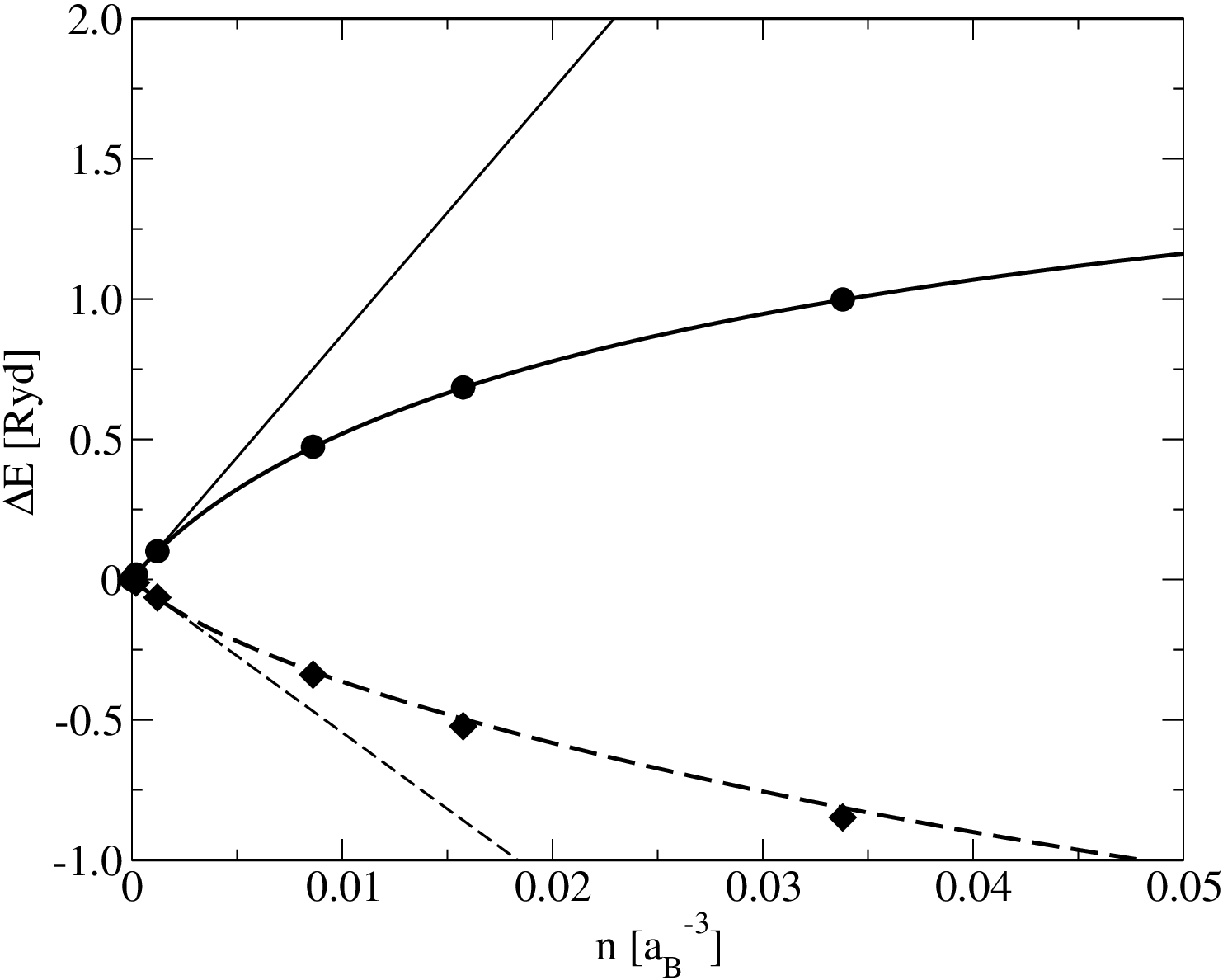}
 \caption{ \label{denshi38a}
 Results of a numerical evaluation of the Pauli shift (dots) and the 
 Fock shift (diamonds) for the temperature $T=5\,000$ K in comparison to the 
 interpolation formulae (\ref{Pauli}) and (\ref{Fock}) (bold lines) and the 
 Boltzmann approximation (thin lines).}
\end{figure}

According to our estimate, the effective binding energy would
disappear at $n_e \simeq 10^{-2}$. This seems to be too early; the
reason is that perturbation approximations tend to overestimate
effects. Better results may be obtained by variational
approximations to the solution of the effective Schr\"odinger
equation. To apply the Ritz variational approach, we have to
symmetrize the Hamiltonian in the effective wave equation
introducing the function $\Psi_n(p) = \psi_n(p)
[1-f_e(p)]^{-1/2}$. We will consider the zero temperature case and
use the {\it ansatz} corresponding to a variable Bohr radius,
\begin{eqnarray}
\Psi_0(p; \alpha) = 8 \pi^{1/2} \alpha^{-3/2} (1 + p^2/\alpha^2)^{-2}.
\label{Psi0}
\end{eqnarray}
Here $\alpha$ is a parameter which characterizes the occupation in the 
momentum space. 
In a more refined approach we may take into account that no states below the 
Fermi momentum are available.
The shift of the binding energy as function of the density is shown in 
Fig. \ref{effener0}.
Within a better approximation we take the Fermi function in the zero 
temperature limit, $f_e(p) = \Theta(p_F-p)$, and evaluate the Pauli blocking 
shift integrating over the wave function $\Psi_0(p; \alpha)$. We obtain
\begin{eqnarray}
\Delta E_1^{\rm Pauli} &\approx&
\frac{4 \pi}{(2 \pi)^3} \int_0^{p_F} d p p^2 [p^2 / \alpha^2 + 1] \psi_0^2 (p)
\nonumber \\ 
&=&\frac{4 \alpha^2}{\pi}\left[\frac{ (p_F^2/\alpha^2 -1)p_F/\alpha}
 {(1+p_F^2/\alpha^2 )^2}+ \arctan (p_F/\alpha )\right].
\label{Pauli0}
\end{eqnarray}
We calculate the temperature dependence of the Pauli blocking term by 
replacing the zero-temperature Fermi function in the interaction term by the 
finite temperature distribution. 
It can be seen that the temperature dependence of the Pauli blocking term 
becomes weak.
Even within the variational approach, the densities where the energy levels 
disappear and consequently full ionization occurs, are evidently still too low 
to explain the observed effects.
We include now the Fock term which is of the same order as the Pauli blocking 
term.
Even if the Fock term is not of primary importance for the disappearance of 
the bound state energy, it has to be included in the total shift of bound and 
scattering states to be consistent (so-called conserving approximations).
In the zero temperature limit we get after some 
transformations the integral
\begin{equation}
\hspace{-1cm}
\Delta E_1^{\rm Fock}=
- \frac{64}{\pi^2}\int_0^{\infty} \frac{p\,d p}{(1+p^2)^4} 
\int_0^{p_F} k\, dk \ln \frac{(p+k)}{|p-k|} 
= -\frac{4}{3 \pi} p_F^3 \frac{5+3 p_F^2}{(1+p_F^2)^2},
\label{Fockasy}
\end{equation}
which reproduces in the low-density limit the value $ (- 20 \pi n_e)$ given 
above in Eq. (\ref{FockB}).
We may estimate the temperature dependence as above in perturbation theory.
For convenience of the numerical procedure in the later variational 
calculations of the free energy we constructed an interpolation formula 
between  the Boltzmann and the zero temperature limits, i.e. Eqs. (6) and (11),
taking into account a few points which we have evaluated numerically,
\begin{equation}
\label{Pauli}
\Delta E_1^{\rm Pauli}= \frac{4}{\pi}
\left\{\frac{p_F[c(T)p_F^2-1]}{[1+c(T)p_F^2](1+p_F^2)}+ \arctan (p_F)\right\}~.
\end{equation}
This is nearly identical to the asymptotic representation except that we
had to introduce a fit function $c(T) = (G(T) - 1)/3$
in order to provide the correct derivative at small densities.
A similar interpolation can be given for the Fock term
\begin{eqnarray}
\label{Fock}
\Delta E_1^{\rm Fock}=-\frac{20\pi}{g}\ln\left(1+g n_e+k n_e^2+l n_e^3\right)
\end{eqnarray}
with the fit parameters $g = 261.65$, $k = 60\,000$, $l = 334\,369$. 
A comparison of the density dependence according to the interpolations 
introduced above with numerical estimates of the integrals is shown in 
Fig. \ref{denshi38a} for $T =$ 5\,000 K. 
The agreement with the data is reasonable for this temperature. 
 In the region of interest
5\,000 K $  <  T  < $ 15\,000 K, the temperature dependence is quite weak. 
The remaining shifts are smaller and will be neglected here, see also \cite{RKKK78,cmf,Arndt96,RedmerReport}.

\section{Ionization equilibrium and thermodynamics in the chemical picture}

We construct the thermodynamic functions of hydrogen by using a
chemical approach to the free energy which recently was applied to
temperatures between $2\,000$ and $10\,000$ K
\cite{befjnrr,bunker,StEb98,FVT,befjrr99,befjrr01,braunes-buch}.
The effects of pressure dissociation, H$_2\rightleftharpoons$ 2H,
and ionization, H $\rightleftharpoons$ e + p, are taken into
account so that the transition from a molecular fluid at low
temperatures and pressures through a partly dissociated, warm
fluid at medium temperatures of some thousand Kelvin to a fully
ionized, hot plasma above $10\,000$ K can be explained.

\begin{figure}
 \includegraphics[width=0.5\textwidth,height=0.4\textwidth,angle=0]{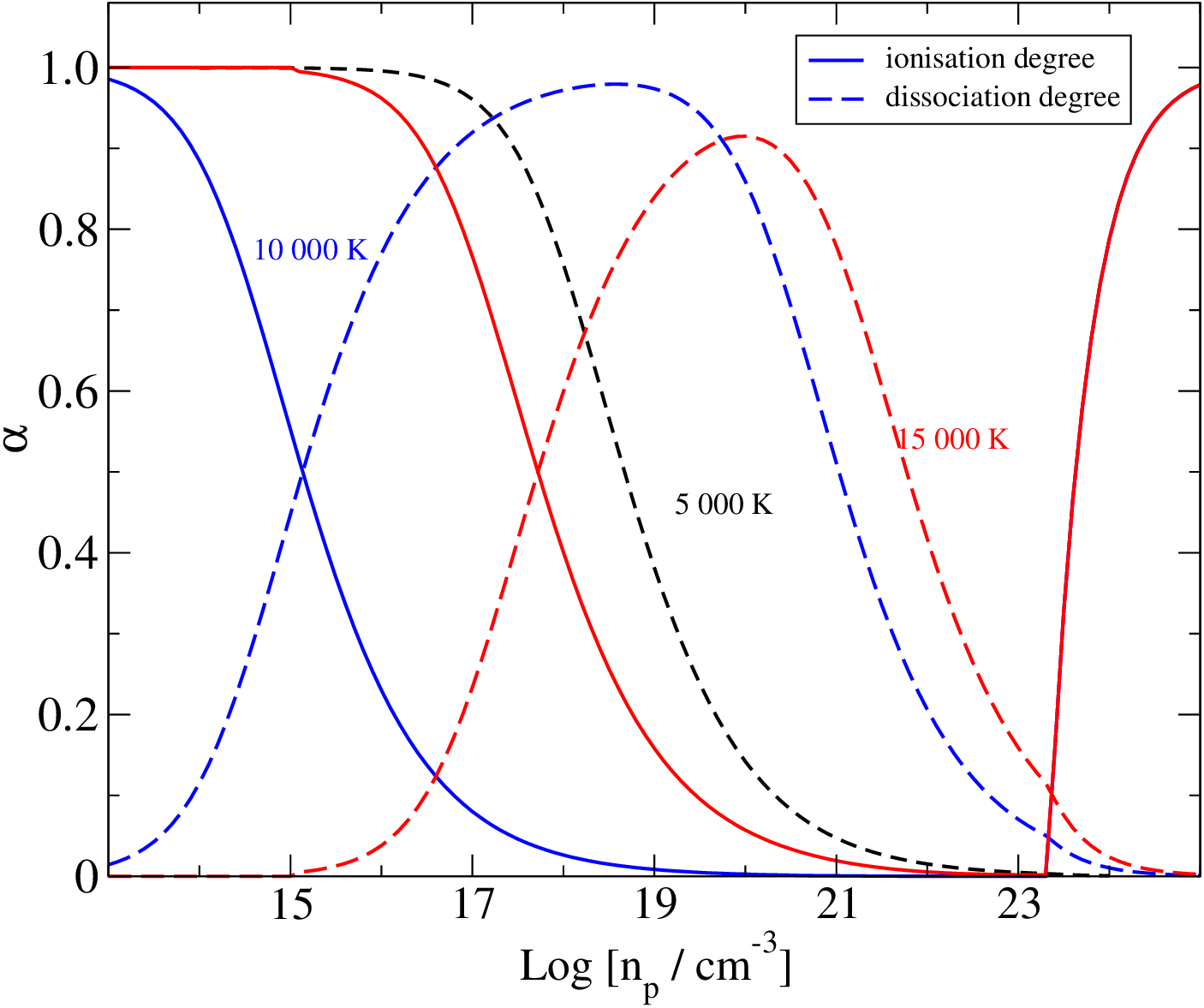}
\caption{ \label{fig:IDDD}
 Degree of ionization and degree of dissociation at $T= 5\,000, 10\,000$ K
 and $15\,000$ K as a function of the total density of protons.}
 \end{figure}

We will not go into the details of the free energy expression used earlier.
Here we take the expression $F_{\rm H}$ as applied so far  \cite{befjnrr,befjrr99,EbHaJRR5} and add
the contribution due to the energy shifts
\begin{equation}\label{F}
F(V,T,{N}) = V (n_{a} + 2 n_{m}) \Delta E(n',T) + F_{\rm H},
\end{equation}
where $n_{a}$ and $n_{m}$ are the number densities of the free
atoms and molecules. The shift of the atomic levels is
approximated by the sum of Pauli and Fock  terms. 
Furthermore, we assumed that molecules are simply composites of two atoms
so that the shifts are additive. 
We took into account ionisation and dissociation processes. 
The degrees of ionisation and dissociation defined by \cite{befjnrr}
\begin{equation}
\alpha =\frac{n_{i}}{n_{i}+n_{a}+2n_{m}}~,~~
\beta_a = \frac{n_a}{n_i + n_a + 2 n_m}~,~~
\beta_m = \frac{2n_m}{n_i + n_a + 2n_m}~,~~
\end{equation}
are the variational parameters of our problem. 
The free energy has to be minimized with respect to them.
We note that $\beta_a$ is the relative amount of protons bound in atoms and 
$\beta_m$ the relative amount bound in molecules. 
Due to the balance relation for the total proton density we have the simplex 
relation $\alpha + \beta_a + \beta_m =1$.
Atoms appear only in a rather narrow region of the temperature - density plane.
The density dependence of the degrees is represented in Fig. \ref{fig:IDDD}.
\begin{figure}
\begin{tabular}{cc}
\includegraphics[width=0.45\textwidth,angle=0]{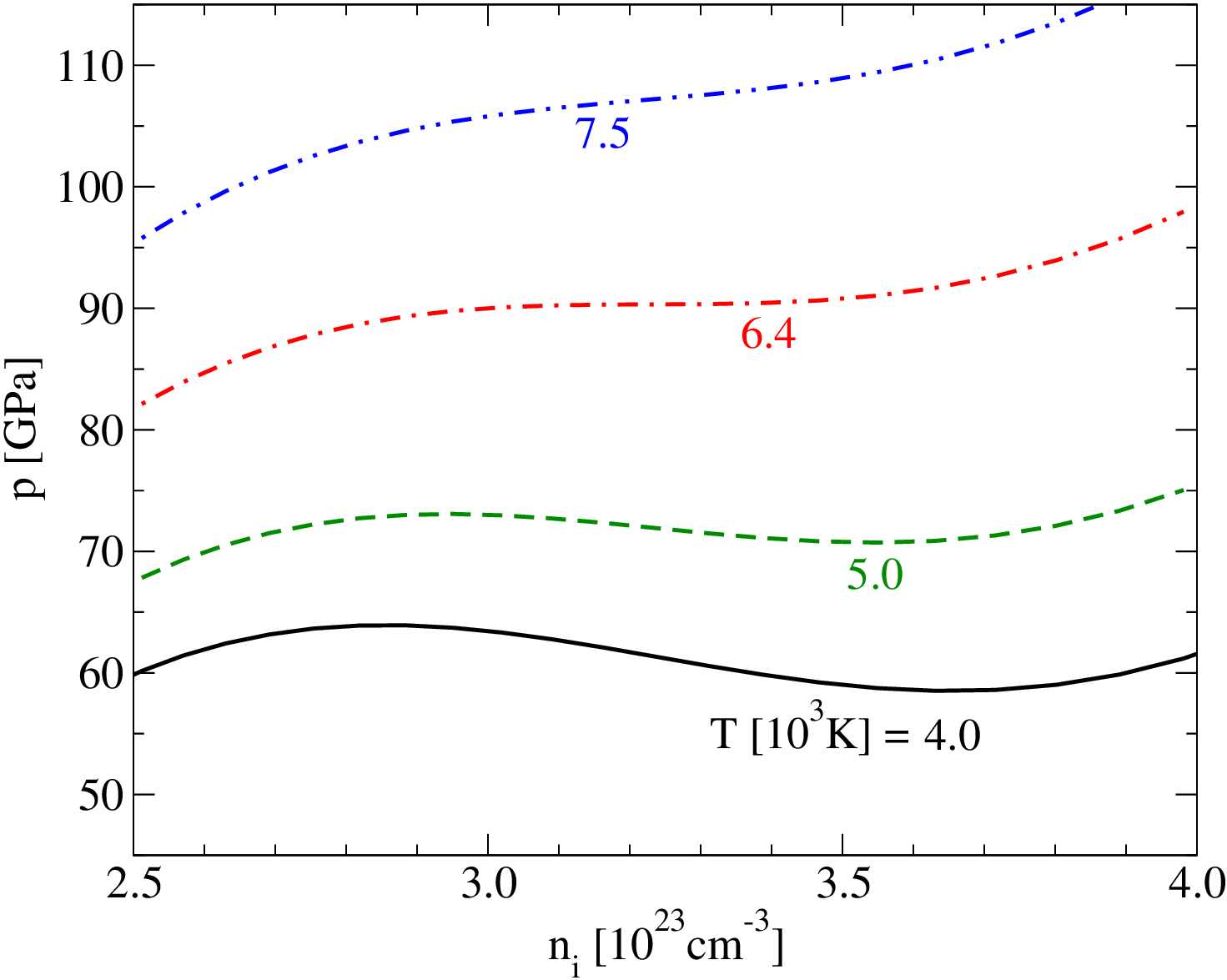}&
\includegraphics[width=0.45\textwidth,angle=0]{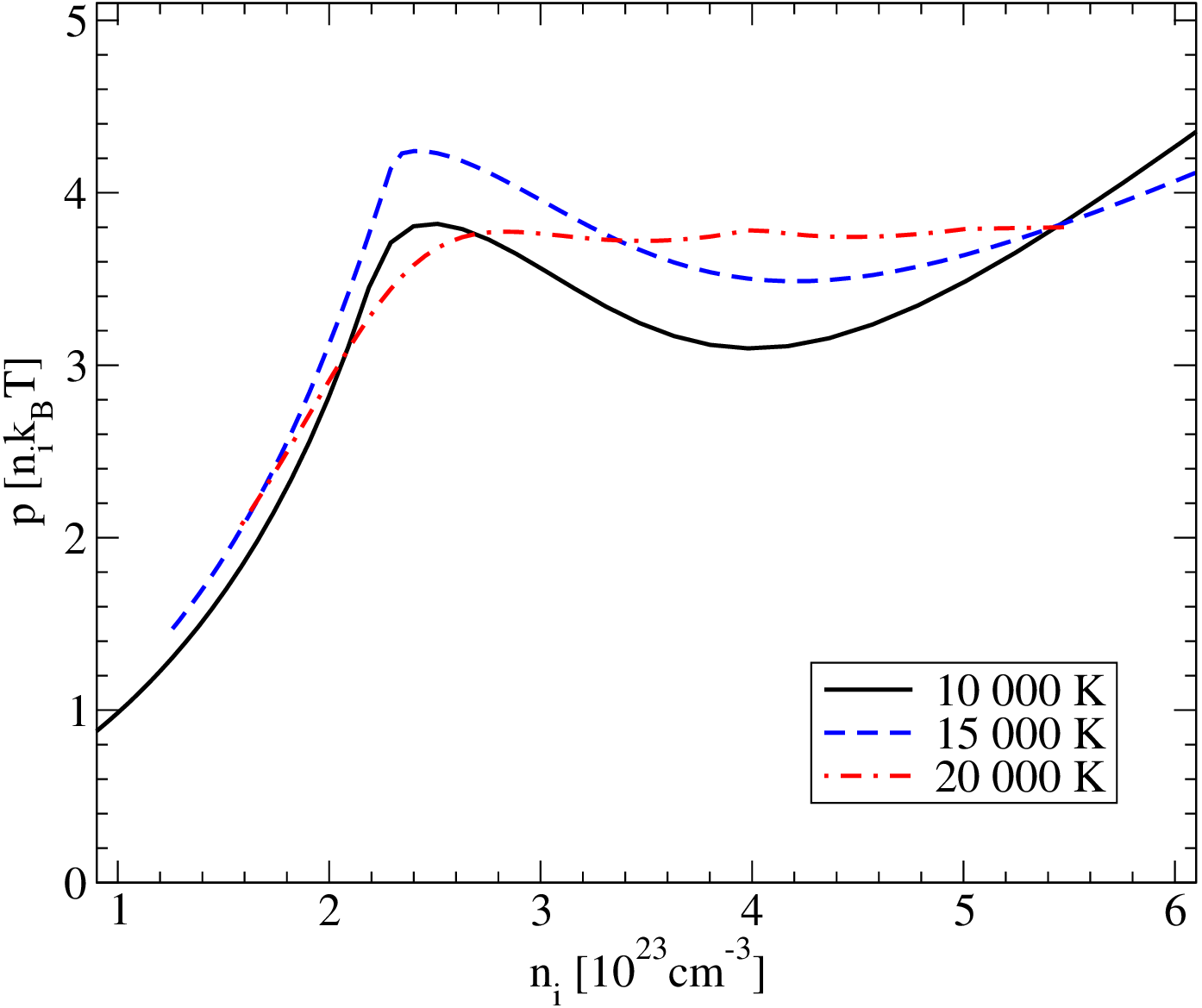}
\end{tabular}
\caption{
Pressure isotherms  (left) and isotherms of the relative pressure  
(right).
\label{PrRel}}
\end{figure}
We calculated the chemical equilibrium by means of a numerical variational 
procedure based on direct minimization of the free energy \cite{EbHaSp03}.
We prefer here the minimization of the free energy as compared to the 
Saha approach because it finds all existing minima, including those at the 
boundaries.
The transition density is between $10^{23}$ and $10^{24}$ protons/cm$^{3}$.
This means we are in the region around 0.5 g/cm$^3$ for hydrogen and
1 g/cm$^{3}$ for deuterium.
The transition to full ionization is rather soft.
In principle, all thermodynamic functions may be calculated from the
free energy (\ref{F}) by derivatives.
Examples of our results for the pressure and the relative pressure 
(in relation to the reference pressure of a fully ionized plasma) are shown 
in Fig. \ref{PrRel}.
We see a phase transition of first order below $6\,450$ K and a diffusive
transition, defined by wiggles in the relative pressure $p/(n_p k_B T)$
up to $20\,000$ K.

\section{Discussion and Conclusions}

In order to understand the transition to
metal-like conductivity in hydrogen at high pressures, we studied the role
of Pauli blocking effects.
We calculated the Pauli and Fock energy shifts by solving effective Schr\"odinger equations for strongly correlated systems.
The ionization and dissociation equilibria were
treated within a chemical approach. We have shown that Pauli blocking effects have a strong influence on
the ionization equilibria and the character of the transition  in the high pressure region.
  We presented explicit calculations of the ionization and dissociation
  equilibria from low to high densities in the
  region 4\,000 K $< T <$ 20\,000 K. An even more detailed description should take into account higher order
contributions to the shifts \cite{Arndt96}. 

The transitions to highly conducting states
occur at densities around $3 \cdot 10^{23}$ protons/cm$^3$.
The corresponding pressures are in the region
of $(0.8 - 1.2) \cdot 10^{11}$ Pa, i.e. around a Mbar.
The first order phase transitions are softer than observed in earlier work.
In the new theory, the first order transition
appears only at $T < $ 6\,450 K, however a diffuse phase transition
detected by wiggles of the relative pressure
remains up to 20\,000  K.

We acknowledge helpful discussions with Wolf D. Kraeft and financial support from the
DFG funded  SFB 652.

\end{document}